\documentclass[twocolumn,showpacs,preprintnumbers,amsmath,amssymb,prl]{revtex4}

\usepackage[latin1]{inputenc}
\usepackage[T1]{fontenc}
\usepackage{ae,aecompl}

\usepackage{graphicx}

\newcommand{\beqn}{\begin{eqnarray}}
\newcommand{\eeqn}{\end{eqnarray}}

\newcommand{\Rfg}[1]{Fig. \ref{F#1}}

\begin{document}

\title{Algebraic statistics of Poincar\'e recurrences in DNA molecule}

\author{Alexey K. Mazur}
%\email{alexey@ibpc.fr}
\affiliation{UPR9080 CNRS, Universit\'e Paris Diderot, Sorbonne Paris Cit\'e,\\
Institut de Biologie Physico-Chimique,\\
13, rue Pierre et Marie Curie, Paris, 75005, France}

\author{D. L. Shepelyansky}
%\homepage[]{http://www.quantware.ups-tlse.fr}
\affiliation{\mbox{Laboratoire de Physique Th\'eorique du CNRS (IRSAMC),
Universit\'e de Toulouse, UPS, F-31062 Toulouse, France}}

%\date{\today}
%\date{August 6, 2015}

\begin{abstract}
Statistics of Poincar\'e recurrences is studied for the base-pair
breathing dynamics of an all-atom DNA molecule in realistic aqueous
environment with thousands of degrees of freedom. It is found that at
least over five decades in time the decay of recurrences is described
by an algebraic law with the Poincar\'e exponent close to $\beta=1.2$.
This value is directly related to the correlation decay exponent $\nu
= \beta -1$, which is close to $\nu\approx 0.15$ observed in the time
resolved Stokes shift experiments.  By applying the virial theorem we
analyse the chaotic dynamics in polynomial potentials and demonstrate
analytically that exponent $\beta=1.2$ is obtained assuming the
dominance of dipole-dipole interactions in the relevant DNA dynamics.
Molecular dynamics simulations also reveal the presence of strong low
frequency noise with the exponent $\eta=1.6$. We trace parallels with
the chaotic dynamics of symplectic maps with a few degrees of freedom
characterized by the Poincar\'e exponent $\beta \sim 1.5$.
\end{abstract}%==========================================

\maketitle

The celebrated Poincar\'e recurrence theorem of 1890 \cite{Poincare:1890} 
guarantees that a dynamical trajectory with a fixed energy 
and bounded phase space will always return
in a close vicinity of the initial state. For dynamical systems
with hard chaos the statistics of Poincar\'e recurrences, and the
related probability to stay in a bounded phase space region behave
similarly to coin flipping and drop exponentially with the return time
$\tau$ \cite{arnold:1968,cornfeld:1982}.  However, in the generic case
of chaos with divided phase space, when islands of integrable motion
are embedded in a chaotic sea \cite{chirikov:1979,lichtenberg:1992},
it was established that the probability distribution of recurrences
$P(\tau)$ is described by an algebraic decay
\begin{equation}
\label{eq1}
P(\tau) \propto 1/\tau^{\beta} \;\; ,
\end{equation}
with the Poincar\'e exponent $\beta \sim 1.5$
\cite{chirikov:1981,karney:1983,chirikov:1984,meiss:1985,chirikov:1999,cristadoro:2008}.
This slow decay originates from sticking of dynamical
trajectories in the vicinity of stability islands and results in a
slow decay of the corresponding atuocorrelation function $C(\tau)
\propto \tau P(\tau)$ \cite{chirikov:1999}. Most of the studies
of the Algebraic Statistics of Poincar\'e Recurrences (ASPR)
considered 2D symplectic maps, notably, the Chirikov standard map,
with recurrence times changing by more than $10$ orders of magnitude
\cite{frahm:2013}. A few recent studies of Hamiltonian systems with a
larger number of degrees of freedom also revealed an algebraic decay
of recurrences with similar values of the Poincar\'e exponent $\beta
\sim 1.3 - 1.5$
\cite{ding:1990,altmann:2007,shepelyansky:2010,altmann:2013}.

The generic nature of the ASPR phenomenon is well established.
It is known to occur on a huge range of physical scales from
electron trajectories for microwave ionization of Rydberg atoms
\cite{buchleitner:1995,benenti:2000} to comet orbits in the Solar
System \cite{shevchenko:2010}. One should expect that it is also
inherent in conformational dynamics of macromolecules. These systems
are characterized by complex energy landscapes, with numerous barriers
and saddle points crossed during thermal motion. It is tacitly assumed
that such dynamics results in a developed chaos with multi-exponential
relaxation decay. Recent experimental evidences suggest, however, that
the ASPR may play an important role, notably, in behavior of the
double helical DNA. The power-law relaxation in the B-DNA  double
helix was discovered \cite{Brauns:02,Andreatta:05} and carefully
studied during the last decade by using the Time Resolved Stokes Shift
(TRSS) experiments
\cite{Gearheart:03,Somoza:04,Sen:05,Sen:06,Andreatta:06,Berg:08}.  In
the last years these results were analyzed with different theoretical
approaches
\cite{Kalosakas:06,Pal:06a,Sen:09,Pal:10,Furse:10a,Furse:10b,Furse:11},
but nevertheless, the possible underlying molecular mechanism remains
elusive. By analogy with the long known power-law kinetics in proteins
\cite{Frauenfelder:88}, this effect in DNA is interpreted in terms of
models developed earlier for glassy systems, with multiple substates,
hierarchical relaxation, mode coupling, etc.  However, unlike
proteins, the B-DNA molecule has only a few well-studied
conformational substates with relatively fast and spatially localized
dynamics that does not resemble those in glasses.  With hydration water
and surrounding ions included, the system becomes more complex, but,
in spite of all efforts, there is no agreement even on whether the
power-law relaxation in the sub-microsecond time range is due to DNA
itself or the hydration water, or both
\cite{Berg:08,Halle:09,Furse:10a}.

We asked, what if, instead of the spin-glass like effect of multiple
degrees of freedom, the power-law relaxation in DNA represents a
manifestation of the ASPR phenomenon. According to experiment, the
decay of the TRSS signal in DNA is described by time
autocorrelation function $C(\tau) \propto 1/\tau^\nu$ with 
$\nu \approx 0.15$ \cite{Andreatta:05}, 
that is, rather close to ASPR in systems
with a few degrees of freedom. This decay is observed over six decades
in time from 10$^{-13}$ to 10$^{-7}$ sec, therefore, one can hope to
detect it by means of all-atom Molecular Dynamics (MD) simulations.
The TRSS signal is obtained by substituting a polarity sensitive dye
coumarine for one of the stacked bases, and the measured effect can be
due to any motion that changes the dye local electric field
\cite{Brauns:02}.  Since in MD different motions are coupled we
started from a trial search of relevant parameters. The statistics of
Poincar\'e recurrences for kinetic energies of selected atoms revealed
only exponential decay. A similar behavior, except for minor details,
was found for conformational transitions of backbone torsion angles.
After a few unsuccessful trials, however, the ASPR has been revealed in
the base-pair breathing motion.

\begin{figure}[ht]
\centerline{\includegraphics[width=0.48\textwidth]{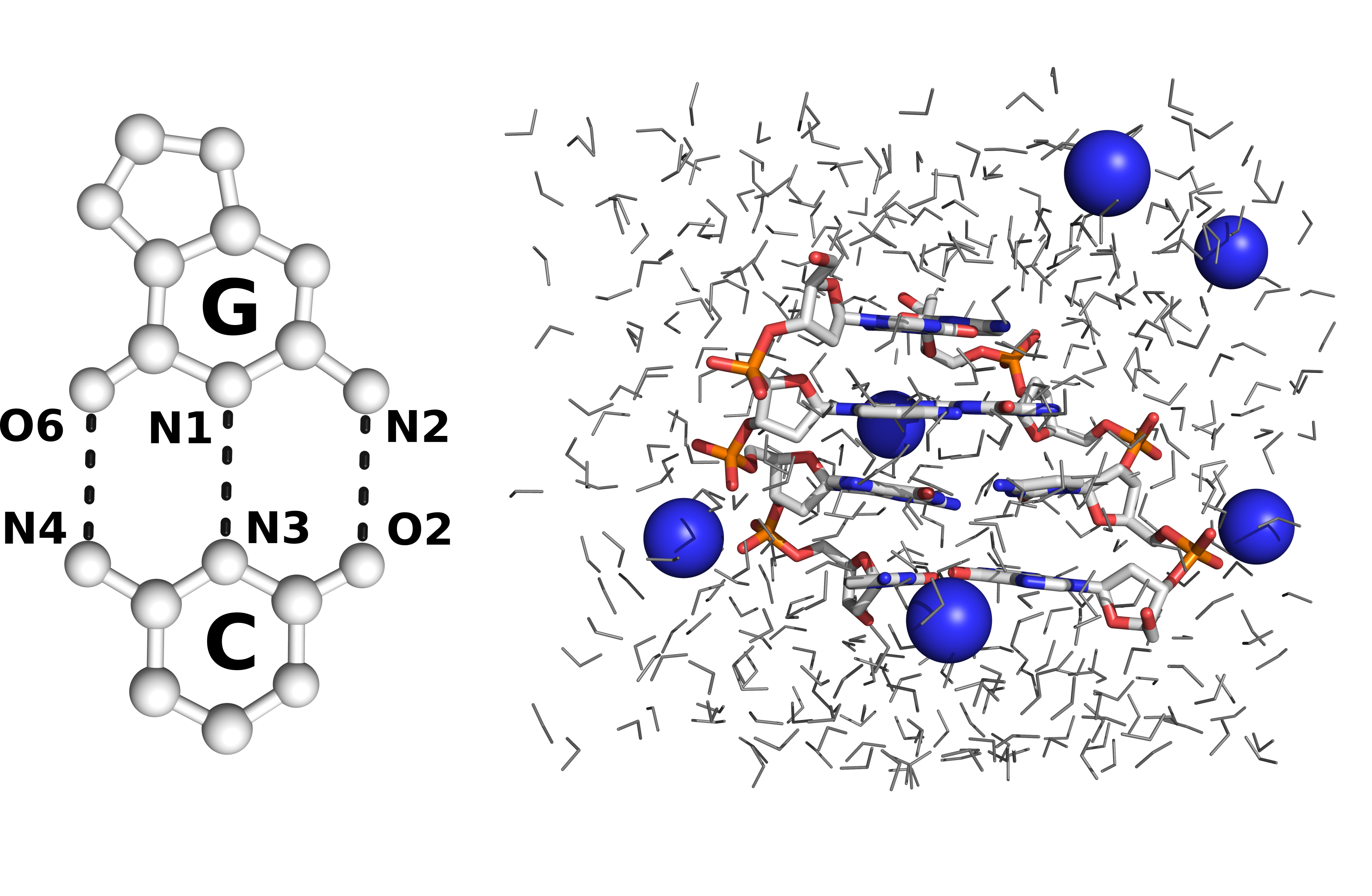}}
\vglue -0.3cm
\caption{\label{Fmol1}(Color online)
The left panel shows the Watson-Crick base pair formed by guanine and
cytosine. The three H-bonds are shown by thick dashed lines.
Their lengths are about 3 \AA. The right panel shows a snapshot of the
model system taken at the end of one of the 65 MD trajectories
involved in statistical analysis. A tetramer fragment of a B-DNA
double helix was built from two identical strands with the
self-complementary base pair sequence GCGC. The six sodium
ions necessary for charge neutralization are shown as spheres.  It is
seen that the four-level stack of base pairs shown in the left panel
remains stable during simulations.}
\end{figure}

The base-pair breathing occurs due to temporal breaking of one or more
hydrogen bonds (H-bonds) in a Watson-Crick (WC) pair. The statistics of
Poincar\'e recurrences was studied for the model system shown in
\Rfg{mol1}. On the left, a GC pair is displayed with three
H-bonds. The right panel shows a tetramer duplex formed by two strands
with identical GC-alternating sequences. This is a minimal symmetrical
structure with two external and two internal base pairs.  The duplex
is placed in a small water box of 489 water molecules with periodic
boundaries. Six sodium ions are added for neutralization. All-atom MD
simulations were carried out in internal coordinates, with fixed
backbone bond lengths and rigid bases, using Hamiltonian equations
\cite{Mzjcc:97} and a symplectic integrator \cite{Mzjchp:99} with the
time step of 0.01 ps \cite{Mzjpc:98}. The recent version of the AMBER
force field \cite{Cornell:95,Perez:07a,Joung:08} was used with SPC/E
water \cite{Berendsen:87}. The system had 3226 degrees of freedom for
1705 atoms.

The base-pair breathing was followed by measuring the distances ($R$)
between the H-bond forming atoms shown in \Rfg{mol1} at every
time step. The stopwatch was started when a given distance exceeded a
certain threshold ($R_{th}$) and stopped once the boundary was crossed in
the opposite direction. These events are called Poincar\'e
recurrences.  The integrated probability distribution $P(\tau)$ is
obtained by counting the number of recurrences with duration larger
than $\tau$ and normalizing it by the total number of events, that is,
$P(0)=1$ by construction.  Function $P(\tau)$ is a very powerful
instrument of analysis because it is positive definite and, due to
statistical averaging over a large number of crossings, stable with
respect to fluctuations (see e.g.  discussion in
\cite{chirikov:1999}).  Importantly, these computations are trivially
parallelizable, that is, the $P(\tau)$ statistics can be accumulated in a
large number of independent MD trajectories. Most of the results
discussed below were obtained by using parallel computations on 65
cores.

\begin{figure}[ht]
\centerline{\includegraphics[width=0.48\textwidth]{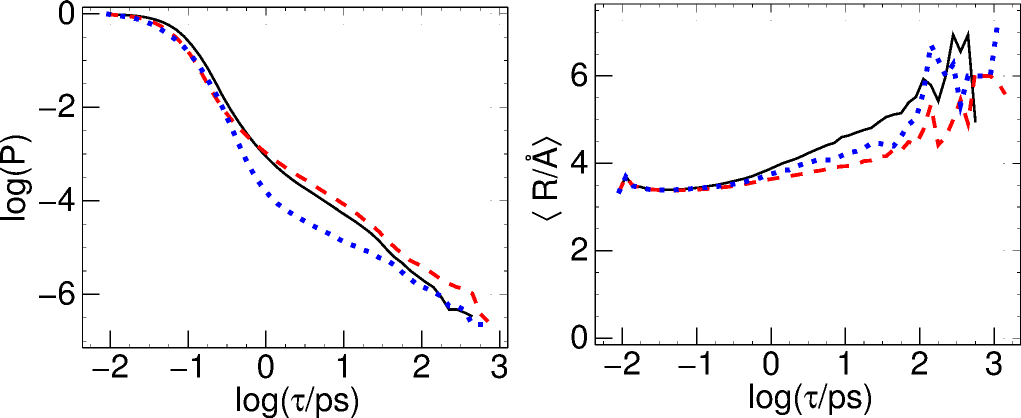}}
\vglue -0.3cm
\caption{\label{F3hb}(Color online) 
The left panel shows statistics of Poincar\'e recurrences $P(\tau)$ for
the three Watson-Crick H-bonds shown in \Rfg{mol1}. The results
for distances O6N4, N1N3, and N2O2 are displayed by solid black,
dashed red and dotted blue curves, respectively. The threshold
distance is $R_{th}= 3.15$ \AA\ in all three cases. The right panel shows
the corresponding dependencies of average bond distances $\langle
R\rangle$ obtained for recurrences of different duration.  Here and in
other figures the logarithms are decimal.}
\end{figure}%=================================================

Representative results obtained for the three H-bonds in
terminal base pairs are shown in \Rfg{3hb}. The left panel displays
double-logarithmic plots of $P(\tau)$ distributions obtained with
$R_{th}$=3.15 \AA. This threshold is close to the equilibrium H-bond
lengths, therefore, a large fraction of recurrences result from
oscillations within the bonded ground state. These motions give for
$\tau$<0.3 ps a characteristic fall of $P(\tau)$ typical for exponential
decays. The right panel of \Rfg{3hb} shows the plots of average
distances for recurrences of different duration. For returns shorter
than 0.3 ps these values remain around 3.5 \AA, that is, the H-bonds
are not broken. With $\tau$>0.3 ps the average distances grow with
$\tau$ and the $P(\tau)$ decay becomes algebraic. The short time boundary
of the power-law relaxation is close to that in TRSS experiments \cite{Andreatta:05}.
Moreover, the decay exponent $\beta \sim 1$ for all three
H-bonds is not far from experimental $\beta -1=\nu\approx 0.15$. To
refine these results we decided to concentrate upon the O6N4 H-bond in
terminal base pairs.  This H-bond is broken easier than other,
therefore, it is an adequate indicator of partial base-pair openings.
The internal base pairs were also checked, but they are opened rarely
and only traces of algebraic decay were detectable in the tails of
$P(\tau)$ distributions.

\begin{figure}[ht]
\centerline{\includegraphics[width=0.48\textwidth]{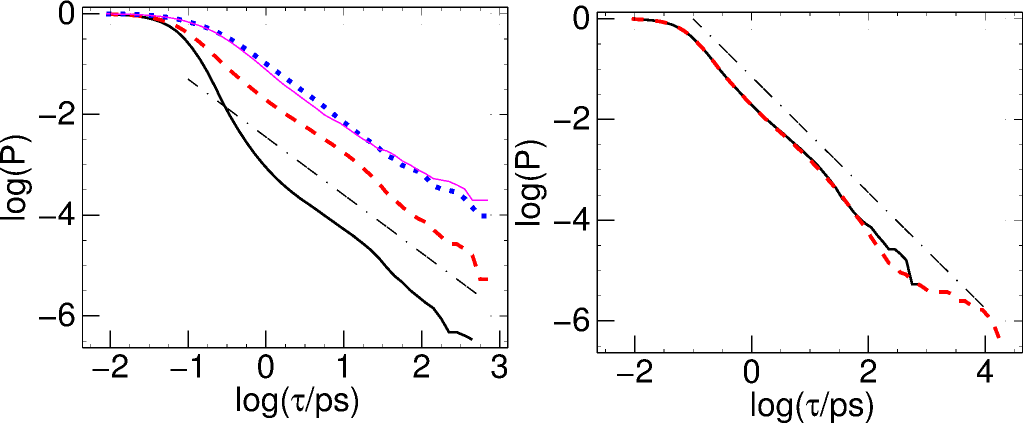}}
\vglue -0.3cm
\caption{\label{Fo6n4}(Color online) 
The left panel shows statistics of Poincar\'e recurrences $P(\tau)$ for
H-bond O6N4 measured at four different threshold distances
$R_{th}$.  The results for $R_{th}$=3.15, 3.55, 4.15, and 4.55 \AA\ are
shown by the solid black, dashed red, dotted blue, and thin solid
magenta curves, respectively. The right panel shows similar statistics
for the threshold of 3.55 \AA\ evaluated in two separate runs. The
solid black curve corresponds to the original simulation (dashed red
curve in the left panel). The second run, shown by dashed red curve,
is carried out with a flat-bottom restraint that prevented the O6N4
distance to go below 3.15 \AA\ thus pushing the system to better
sampling of long returns. 
In both panels the dash-and-dot straight line shows the power law
decay with the Poincar\'e exponent $\beta = 1.15$.}
\end{figure}%==============================================

To get a better estimate of the short-time boundary of the algebraic
decay in $P(\tau)$ we need to remove the contribution of quick returns
that do not result in H-bond opening. To this end the threshold $R_{th}$
was gradually increased up to 4.55 \AA. The results of these
computations are displayed in the left panel of \Rfg{o6n4}. It is seen
that the short time hump is essentially removed already with
$R_{th}$=3.55 and that for the true base-pair breathing the algebraic
decay starts from very short recurrences of only 0.1 ps.  This refined
boundary is in excellent agreement with the experimental data
\cite{Andreatta:05}.

These computations were also used to refine the estimate of the
exponent $\beta$. Linear fits of the plots in the left panel of
\Rfg{o6n4} were carried out in the range $-0.5 \leq \log (\tau/ps)
\leq 2.5$, which gives exponent values $\beta$=1.27, 1.20, 1.13, and
1.05 for thresholds $R_{th}$=3.15, 3.55, 4.15, and 4.55 \AA,
respectively. It is seen that there is a moderate influence of
$R_{th}$ upon the exponent, which can be related to the fact that
longer recurrences correspond to larger atom-atom separations
(\Rfg{3hb}). Note that the $\beta$ values obtained give the exponent
of correlation decay $\nu=\beta - 1$ very close to the experimental
TRSS value ($\nu\approx 0.15$) \cite{Andreatta:05}. For visual
comparison, the $P(\tau)$ decay predicted from experiment is shown in
\Rfg{o6n4} by the dash-and-dot lines.

The experimental long-time boundary occurs at $\approx$100 ns and it
corresponds to the maximal resolution of the TRSS method
\cite{Brauns:02,Andreatta:05}. The long-time boundary in our
computations is limited by sampling. We tried to push it somewhat
further by taking into account that H-bonds in our simulations spend
almost all the time oscillating around the ground energy minimum. To
reduce this non-productive time a flat-bottom restraint was added that
prevented the O6N4 distances to go below 3.15 \AA. It is understood
that this simple {\em ad hoc} trick perturbs realistic dynamics, but
the information it provides may be useful. The results of these
computations shown in the right panel of \Rfg{o6n4} confirm that the
sampling is indeed improved, with the time range of the approximate
power law extended by about an order of magnitude. For shorter
duration the $P(\tau)$ distribution reproduces the curve obtained
without the restraint. Therefore, one can reasonably expect that the
long-time boundary of the power-law decay of $P(\tau)$ occurs at least
at $\approx$10 ns.

\begin{figure}[ht]
\begin{center}
\includegraphics[width=0.48\textwidth]{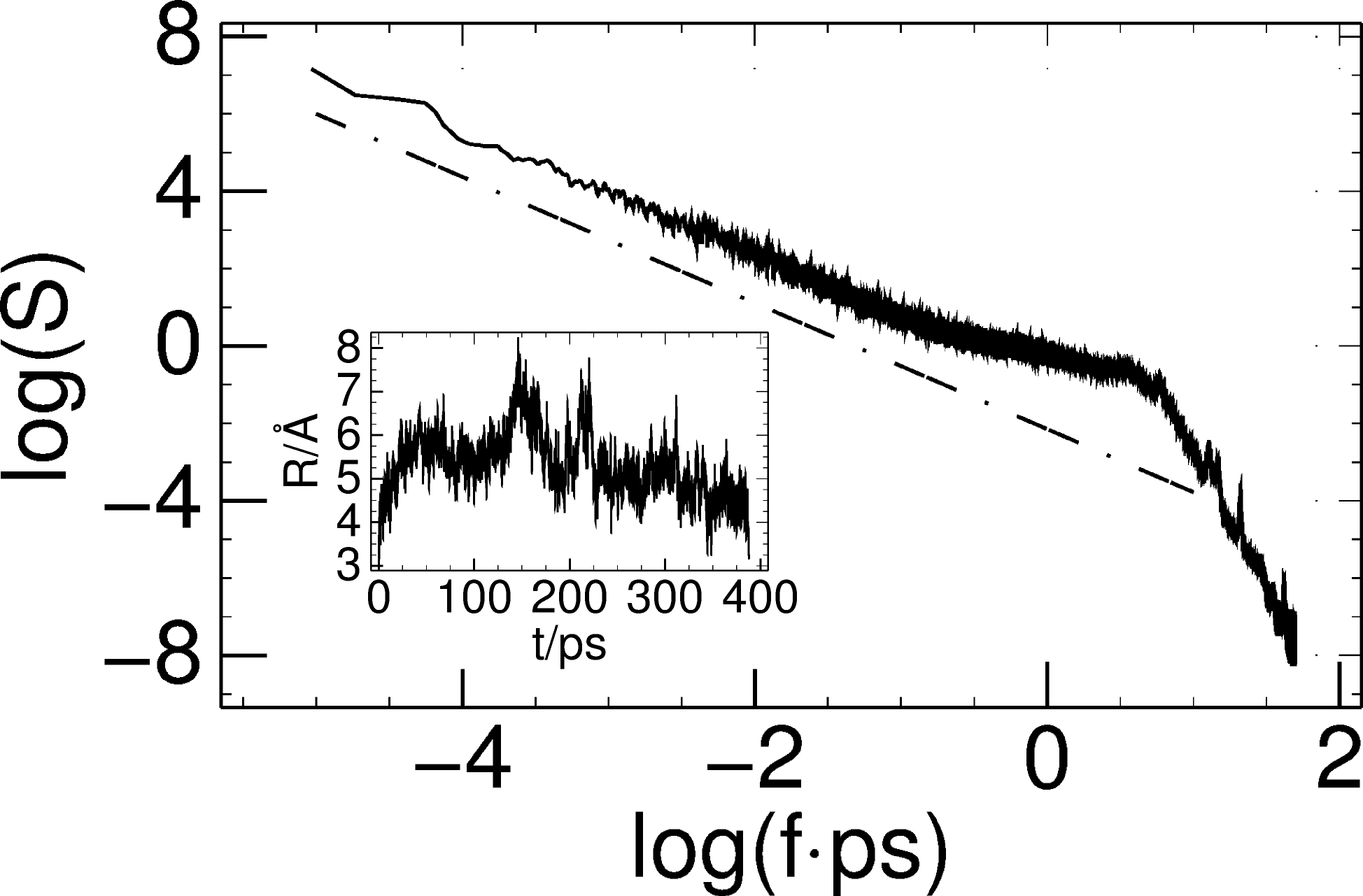}
\end{center}
\vglue -0.3cm
\caption{\label{Fpower}(Color online) 
The average power spectral density of time fluctuations of the O6N4
distance obtained from 16 independent trajectories of 10 ns each, with
the data stored at every time step. The dash-and-dot straight line
shows the power law decay with the exponent $\eta=1.63$ obtained from
the linear fit in the range $-5 < \log (f \cdot ps) < -1$.  The insert
shows a representative time trace of the same parameter between two
consecutive crossings of the threshold distance $R_{th}= 3.15$ \AA.  }
\end{figure}%===============================================

Encouraged by the foregoing findings, notably, the surprisingly good
agreement with experiment obtained with no adjustable parameters, we
decided to study the power spectrum of the fluctuations of the O6N4
distance. The results shown in \Rfg{power} reveal a strong power-law
growth of density $S$ for low frequencies. However, the exponent
$\eta=1.63$ estimated by linear fitting does not seem to be related to
neither experimental data nor the Poincar\'e recurrences studied above.
This paradoxical observation is discussed further below.

To shed light on the possible origin of the foregoing results we use
the virial theorem \cite{landau} for analysis of chaotic dynamics in
systems with representative polynomial potentials. The corresponding
Hamiltonian reads
\begin{equation}
\label{eq2}
H(p,r) =p^2/2 - a/r^m = E \, ,
\end{equation}
where $(p,r)$ is the pair of conjugated variables, with momentum $p$ and
radial coordinate $r$. The energy $E<0$ corresponds to a bounded motion;
$a \sim 1$ and $m >0$ are numerical coefficients (the mass is taken as
unity).  We assume that in addition to radial dynamics there is a
chaotic motion in angle degrees of freedom. For this system the
action can be estimated as $J \sim p r \sim (b E)^{(m-2)/(2m)}$ with a
certain numerical constant $b<0$, which follows from relationships $p^2
\sim 1/r^m$ and $E \sim p^2$.  Thus we have $bE \sim |E|   \sim
J^{2m/(m-2)}$, and $\omega = dH/dJ \sim J^{(m+2)/(m-2)} \sim 1/\tau$,
where $\omega$ is a frequency of motion and $\tau$ is the related
characteristic time scale.

As a result, the measure $\mu$ related to the sticking time scale
$\tau$ is obtained as $\mu \sim J \sim 1/\tau^{(m-2)/(m+2)}$, which
follows from the fact that in Hamiltonian systems the measure is
proportional to the phase volume, that is, $\mu \sim \int J d\theta
\sim J \times 2\pi$, where $\theta$ is the angle variable conjugated
to action $J$.  According to \cite{chirikov:1999} we have $P(\tau)
\sim d \mu/d\tau \sim \mu/\tau$ where  $\mu(\tau)$ is the measure of a
region where a trajectory is stuck for the time $\tau$.  This follows
from the ergodicity relation according to which the measure of a
region is proportional to the time spent by the trajectory in this region
$\mu(\tau) \sim \tau P(\tau)/{\langle \tau \rangle}$, where $\langle
\tau \rangle = \int_0^\infty P(\tau) d\tau$ is the average time of
recurrences \cite{arnold:1968,cornfeld:1982}.  From these relations
one obtains the following expression for the Poincar\'e exponent
\begin{equation}
\label{eq3}
\beta=2m/(m+2) \, .
\end{equation}
Here $P(\tau)$ can be considered as an integrated probability of
Poincar\'e recurrences or as a survival probability in a given region
for time periods longer than $\tau$ since both are proportional to
each other \cite{chirikov:1999,frahm:2013}.

Consider some earlier studied potentials with different $m$.
With $m=1$ we get the Kepler problem. This case appears in the
microwave ionization of Rydberg atoms
\cite{buchleitner:1995,benenti:2000}, and also in the comet
\cite{shevchenko:2010} or dark matter \cite{lages:2013} dynamics in
the Solar System affected by Jupiter. In both cases the energy change
occurs when the particle passes near the perihelion (near the nuclei or
near the Sun). This energy change produces chaotic dynamics in the
system. In this case the measure $\mu \sim J \sim 1/\sqrt{|E|}$ is
diverging at $|E| \rightarrow 0$ and from (\ref{eq3}) we have $\beta
=2/3 <1$, which agrees with the analytical and numerical results
reported in \cite{borgonovi:1998}. For $m=2$ we have a period
independent of action, which gives $\beta=1$ without decay of
correlations, that is, $C \sim \tau P(\tau) \propto const$ with
$\nu=\beta -1 =0$.

For the most relevant case of dipole-dipole interactions that are
dominant in H-bonds and, more generally, in neutral polar systems like
B-DNA with ions in water, we have $m=3$ and $\beta=1.2$. The last value
is close to that obtained in our numerical simulations as well as that
corresponding to exponent $\nu=\beta-1$ found in the TRSS experiments.
We believe that the above estimates correctly capture the main
physical effects in the dynamics of this complex system and are at the
origin of the observed slow algebraic decay of Poincar\'e recurrences.
It is understood that in systems with thousands of atoms like that
studied here other factors can interfere. It is possible that Coulomb
forces ($m=1$) of locally uncompensated charges are responsible for a
certain reduction of $\beta$ to a slightly smaller value compared to
the above theoretical estimate. We also note that for the van der
Waals potential we have $m=6$ with the corresponding $\beta=3/2$.

Finally, consider the result shown in \Rfg{power}. The measure of
sticking regions is
$\mu \sim J \sim |E|^{1/6} \sim 1/\sqrt{r} \sim 1/\tau^{1/5}$. It
decreases with large $\tau$,
but the typical atom-atom separations are growing as $R_b \sim r \sim
\tau^{2/5}$. The correlation function $C(\tau) = \langle r(t+\tau)
r(t) \rangle_t$ is commonly defined for a bounded variable $r(t)$, so
that the average square variation is $ \langle r(\tau)^2 \rangle_t
\sim \tau \int C(\tau) d \tau \sim \tau^{2 - \nu}$, where $\nu = \beta
- 1 < 1$ is the correlation decay exponent related to the
super-diffusive growth. In the present case, we have a variable that
grows with $\tau$, which gives an additional contribution to the
average square variation $ \langle r(\tau)^2 \rangle \sim R_b^2 \tau^2
C(\tau) \sim R_b^2 \tau^{3-\beta} \sim \tau^\kappa$ with $\kappa=3
\beta - 1 =2.6$. Using the Wiener-Khintchine relation (see
e.g.  \cite{robinson:1974}) between the square variation of a time
dependent variable $r(t)$ and its spectral density $S(f)$ we obtain
\begin{equation}
\label{eq4}
S(f) \propto 1/f^\eta \; , \; \eta = \kappa - 1 = 3 \beta - 2 = 1.6
\end{equation}
in good agreement with \Rfg{power}. In other words, we have a very
strong low frequency noise, with the exponent $\eta$ larger than usual
\cite{kogan:1996}, that results from only the inherent chaotic
dynamics of the system, with no external noise involved.

In summary, using all-atom MD simulations, we uncovered the existence
of algebraic decay of Poincar\'e recurrences in the base-pair
breathing motion of the B-DNA molecule, with the decay exponent $\beta
\approx 1.2$ and a strong divergence of spectral density of vibration
motion with the exponent $\eta\approx 1.6$ over at least five decades
in time from 10$^{-13}$ to 10$^{-8}$ sec.  These results are well
described by the proposed theory of Poincar\'e recurrences for chaotic
dynamics in polynomial potentials, assuming a dominant contribution of
dipole-dipole interactions. The theory correctly captures the
qualitative origin of this effect in MD in spite of the difference in
the number of degrees of freedom. The exponent $\beta \approx 1.2$
predicts a power-law relaxation of the corresponding correlations with
the exponent $\nu = \beta -1 \approx 0.2$. This prediction as well as
the time scale of the algebraic decay in MD are in striking similarity
with TRSS experiments, suggesting that this effect is due to the
base-breathing motion. Further studies should help to clarify
the relationship between the inherent dynamical chaos in DNA and these
experimental data.

\bibliography{dnapoincare}

\end{document}